\documentclass[a4paper, 12pt]{iopart}
\usepackage{color, psfrag, epsfig, graphicx, geometry, wrapfig}
\usepackage{mathrsfs}
\usepackage{url}
\usepackage{multirow}

\usepackage{ulem}
\definecolor{red}{rgb}{1,0,0}

\begin{document}

\title[Stellar core-collapse waveform decomposition]{
Rotating stellar core-collapse waveform decomposition: a Principal Component
Analysis approach}

\author{Ik Siong Heng}

\ead{i.heng@physics.gla.ac.uk}

\begin{abstract} \\

This paper introduces the use of Principal Componenent Analysis
as a method to decompose the catalogues of gravitational waveforms 
to produce a set of
orthonormal basis vectors. We apply this method to a set of
gravitational waveforms produced by rotating stellar core-collapse simulations 
and compare the basis vectors obtained with those 
obtained through Gram-Schmidt decomposition. 
We observe that, for the chosen set of waveforms, the performance of the 
two methods are comparable for minimal match requirements up to 0.9,
with 14 Gram-Schmidt basis vectors 
and 12 principal components required for a minimal match of 0.9.
This implies that there are many common features in the chosen waveforms. 
Additionally, we observe the chosen waveforms have very similar
features and a minimal match of 0.7 can be obtained by decomposing 
only waveforms generated from simulations with A=2.
We discuss the implications of this observation and
the advantages of eigen-decomposing waveform catalogues with 
Principal Component Analysis. 
\end{abstract}

\section{Introduction}

The current global network of interferometric detectors
(GEO\,600~\cite{GEO},
LIGO~\cite{LIGO}, TAMA\,300~\cite{TAMA} and
Virgo~\cite{Virgo}) have been scanning the sky for gravitational
wave signals with unprecedented
sensitivities.
The prospect of detection is very real and once a detection is made, we must ask what
can be inferred about the source from the detected gravitational wave signal. 
While this issue must be addressed for all signal types, we choose to focus on core-collapse
supernovae here.
Numerical relativity simulations have predicted several sets of gravitational wave signals
or {\it waveform catalogues} due to rotating core-collapse supernovae
(see~\cite{ott-review} and references therein).
The features of each waveform produced by the simulations vary depending on the physics
employed and the chosen initial parameters.
The predicted waveforms can be used as inputs into parameter estimation algorithms which
can, for example, calculate the likelihood that the detected signal corresponds to one of
the predicted waveforms.
However, from a cursory inspection of the waveform catalogues, one can see that there
are many common features in the predicted waveforms, especially for waveforms in the same
catalogue. 
By decomposing the waveforms into a set of orthonormal basis vectors, we can greatly reduce 
the computation costs of the parameter estimation stage by concentrating on a subset of
basis vectors that encompass the main features of the chosen waveforms.

We propose using Principal Component Analysis (PCA) to create an orthonormal set of basis
vectors. 
Broadly speaking, PCA transforms a correlated, multi-dimensional data set into a set of
orthogonal components.
This is achieved by determining the eigenvectors and eigenvalues of
the covariance matrix of the data set.
The first principal component is the eigenvector with the largest corresponding eigenvalue.
It is a the linear combination of the
original variables which accounts for as much of the variability in the
data as possible. 
Similarly, the second principal component is the linear combination
which accounts for as much of the remaining variability as possible --
subject to the constraint that it is orthogonal to the first principal
component -- and so on.
In recent years, PCA has been applied to a number of astrophysical problems
(see \cite{disney} for a recent example), 
such as spectral
classification, photometric redshift determination and morphological
analysis of galaxy redshift surveys, as well as wider class of image
processing and pattern recognition problems across a range of scientific
applications.
For a detailed account of the
statistical basis of PCA the reader is referred to, for example, Morrison
(1967)~\cite{morrison} or Mardia {\it et al.}
(1979)~\cite{mardia}.

It must be noted that this is not the first approach proposed to
decompose a waveform catalogue.
Brady and Ray-Majumder~\cite{brady} have previously applied Gram-Schmidt decomposition to
the Zwerger-M\"{u}ller~\cite{zm} and Ott {\it et
al.}~\cite{ott2004} waveform catalogues.
Additionally, Summerscales {\it et al.}~\cite{summerscales} developed a Maximum Entropy based method 
to identify the presence of a gravitational wave signal and demonstrated the method's 
ability to extract the correct amplitude and phase of waveforms from a 
catalogue by Ott {\it et al.} (2004)~\cite{ott2004}.

In this article, we choose the waveform catalogue generated by simulations peformed by 
Dimmelmeier {\it et al}~\cite{dimmelmeier07} to demonstrate the use of PCA.
These simulations focus on rotating stellar core-collapse and the subsequent bounce.
While this phase of a supernova has been the focus of numerous simulations over the years,
recent simulations by Burrows {\it et al.} have produced large post-bounce gravitational wave 
signals due to acoustic shock mechanisms~\cite{ott-burrows}.
There is still much discussion about this mechanism so we choose to concentrate only on
the gravitational wave signal produced by the core-collapse phase.
Moreoever, we would like to stress that the choice of catalogue here is, to a large extent, 
arbitrary and the methods discussed
in this paper can be easily applied to the other catalogues or any combination thereof.
Additionally, we compare the basis vectors obtained by PCA and
Gram-Schmidt
decomposition by applying them to the same set of waveforms.
We then describe our observations of the waveform catalogue before discussing our observations.

\section{Methods}

\subsection{Gram-Schmidt decomposition}

Gram-Schmidt (GS) decomposition is a recursive method for decomposing a set of waveforms 
to create a set of orthonormal basis vectors~\cite{golub}. 
It was first applied to a supernova catalogue 
by Brady and Ray-Majumder~\cite{brady}.
For completeness, the main points of this method are reviewed below.

In GS decomposition, one begins by selecting a waveform from the data
set as the first basis vector. To create a second basis vector, the first basis
vector is first projected onto the next waveform to be included into the set of basis
vectors. The projected component is then subtracted from the second waveform and
the resulting vector is orthogonal to the first basis vector. One continues this
process by subtracting the sum of the projections of all exisiting basis vectors
onto the desired waveform.
This is done recursively until the desired number of waveforms are included into the set of basis 
vectors. More explicitly, for a set of waveforms,
$\{H_1,H_2,...,H_M\}$,
the orthonormal basis vectors, $\{e_1,e_2,...,e_M\}$, are
\begin{equation}\label{gs1}
e_i = \bar{H}_i / \sqrt{(\bar{H}_i,\bar{H}_i)},
\end{equation}
with
\begin{equation}\label{subproj}
\bar{H}_i = H_i - \sum^{i-1}_{j=1} (H_i,e_j)e_j
\end{equation}
and $i=1,..,M$ where $M$ is the total number of waveforms.
Here, the brackets denote an inner product. Explicitly, the inner
product for two vectors $a$ and $b$, each of length $n$ is given
by 
\begin{equation}
(a,b) = \sum^n_{i=1} a_i b_i
\end{equation}
where $a_i$ denotes the $i^{\rm th}$ element of the vector $a$.

Note that the second term in Equation~\ref{subproj} is the sum of the projection from all
previously formed basis vectors. 
Therefore, $\bar{H}_i$ is the residual waveform not described by previously generated basis
vectors.

Brady and Majumder point out that the choice of the first waveform is
chosen arbitrarily and may produce a basis vector set that spans the
waveforms most efficiently.
Therefore, the basis set is constructed repeatedly with a different
initial waveform chosen each time until the basis vector that spans
the waveforms' parameter with the fewest number of basis vectors is
obtained.

\subsection{Principal Component Analysis}

In Principal Component Analysis (PCA), a basis set is formed by determining the eigenvectors
of the covariance matrix of the desired data set.
In the context of this article, let us arrange the waveforms from the catalogue
$\{H\}$ into a matrix ${\bf H}$ such that each column corresponds to one of the
waveforms, $H_i$. For $M$ waveforms, each of length $N$, the matrix ${\bf H}$ 
has dimensions of $N \times M$ and the covariance matrix for ${\bf H}$ is
calculated by
\begin{equation}\label{cov_mat}
{\bf C}=\frac{1}{M} {\bf H}{\bf H}^{\rm T}.
\end{equation}
where ${\bf C}$ is the covariance matrix with dimensions $N
\times N$ for waveforms with length $N$.
%
%
The normalised eigenvectors of ${\bf C}$ form a set of basis
vectors, $\{e_1,e_2,...,e_M\}$, that span the
parameter space defined by the waveforms in ${\bf H}$. 
Note that in PCA, the eigenvalues of the covariance matrix tell us how well each
eigenvector spans the parameter space of the waveform catalogue.
The eigenvectors are, therefore, ranked 
by their corresponding eigenvalues, with the first principal component having the largest 
eigenvalue.

Supernovae waveforms have significant energies at high frequencies ($\sim 1$\,kHz), so $N$ 
can be about 1000 data samples at LIGO (16384\,Hz) data sampling
rates or more at Virgo (20\,kHz) sampling rates.
Determining the eigenvectors of a matrix of such dimensions is
computationally expensive.
A common method of avoiding this computationally intensive
operation (see~\cite{horn} for example)
is to first calculate the
eigenvectors, ${\bf v}$, of ${\bf H}^{\rm T}{\bf H}$ such that
\begin{equation}\label{eig_v_1}
{\bf H}^{\rm T}{\bf H}{\bf v}_i = \lambda_i {\bf v}_i,
\end{equation}
where $\lambda_i$ is the corresponding eigenvalue for each
eigenvector. Then, by pre-multiplying both sides by ${\bf H}$, we have
\begin{equation}\label{eig_v_2}
{\bf H}{\bf H}^{\rm T}{\bf H}{\bf v}_i = \lambda_i {\bf H}{\bf v}_i.
\end{equation}
If we rewrite Equation~\ref{cov_mat} so that the covariance matrix takes the form
${\bf C} = {\bf H}{\bf H}^T$,
then ${\bf H}{\bf v}_i$ are the eigenvectors of the covariance
matrix.~\footnote{The method laid out here is similar to
performing
a singular value decomposition (SVD) \cite{SVD} of the matrix
${\bf H}$. In SVD, equations~\ref{eig_v_1} and \ref{eig_v_2} are the
equivalent of using the right-singular vectors, which are
the eigenvectors of ${\bf H}^{\rm T}{\bf H}$, to determine the
left-singular vectors, which are the eigenvalues of ${\bf H}{\bf H}^{\rm T}$.}
So, for $M \ll N$, we can determine the eigenvectors of
covariance matrix by first calculating the eigenvectors of the
smaller ${\bf H}^T{\bf H}$ which is an $M \times M$ matrix, 
thereby significantly reducing computation costs.

\section{Results}

\subsection{The waveforms}


The waveform catalogue used here to demonstrate the use of PCA were
produced by Dimmelmeier {\it et al}~\cite{dimmelmeier07}.
These waveforms are generated from axisymmetric general relativistic
hydrodynamic simulations of stellar core-collapse to a proto-neutron star. 
They use the microphysics equation of state from Shen {\it et
al.}~\cite{shen} with a 20 $M_\odot$ progenitor model from Woosley
{\it et al.}~\cite{woosley}.
There is a total of 54 waveforms in this catalogue generated by models which are
parameterised by initial differential rotation, $A$, and
the ratio of the rotational kinetic to gravitational energies, $\beta$.
The values of $\beta$ are increased from $0.05\%$ to $4\%$ in 18 steps while
three values of initial differential rotation were used. 
The three values, labelled $A=1$, $A=2$ and $A=3$,
corresponding to differential rotation occurring at 50000\,km (almost uniform rotation), 
1000\,km and 500\,km respectively.
According to the rotation law~\cite{rotation_law} the angular velocity
has dropped to 1/2 its central value at distance of A from the rotation
axis. Hence, smaller values of A correspond to more differential rotation. 

\subsection{Comparing GS and PCA basis vectors}\label{sec:comp}

%
We introduce the {\it match} parameter, $\mu$, to quantify how well a set of basis vectors 
reconstructs a specified waveform. 
For a waveform, $H_i$, $\mu_i$ is calculated by 
summing the projections of the desired number of basis vectors, $Z$, onto the waveform
such that
\begin{equation}
\mu_i = \left | \left | \sum^Z_{j=1}{ (H_i , e_j)e_j} \right | \right |
\end{equation}
where $e_j$ are the orthonormal basis vectors determined by the methods described in the
previous section~ref{brady}. As with equations \ref{gs1} and \ref{subproj}, the brackets denote an inner product. If we normalise
the set of waveforms, then $\mu_i$ will be equal to 1 if the sum of the projections
of the basis vectors match at particular waveform, $H_i$, exactly. 

It is clear that $\mu$ will be equal to 1 for all waveforms in
the catalogue if we use all basis vectors decomposed from the
catalogue ($Z = M$).
However, it is interesting to calculate the smallest match
obtained for any waveform in the catalogue (commonly referred to
as {\it minimal match}) if we use a subset of basis vectors. 
The minimal match, $\mu_{\rm min}$, is often used in templated matched
filter searches for signals with well-modelled waveforms.
For such searches, the basis vectors form a bank of templates and
minimal match is used to characterise how well the desired parameter space
is covered by the template bank.
To maximise the detection probability, one would maximise minimal
match with the smallest number of templates so as to minimise
computational time.

\begin{figure}[!h]
\centering
\epsfig{file=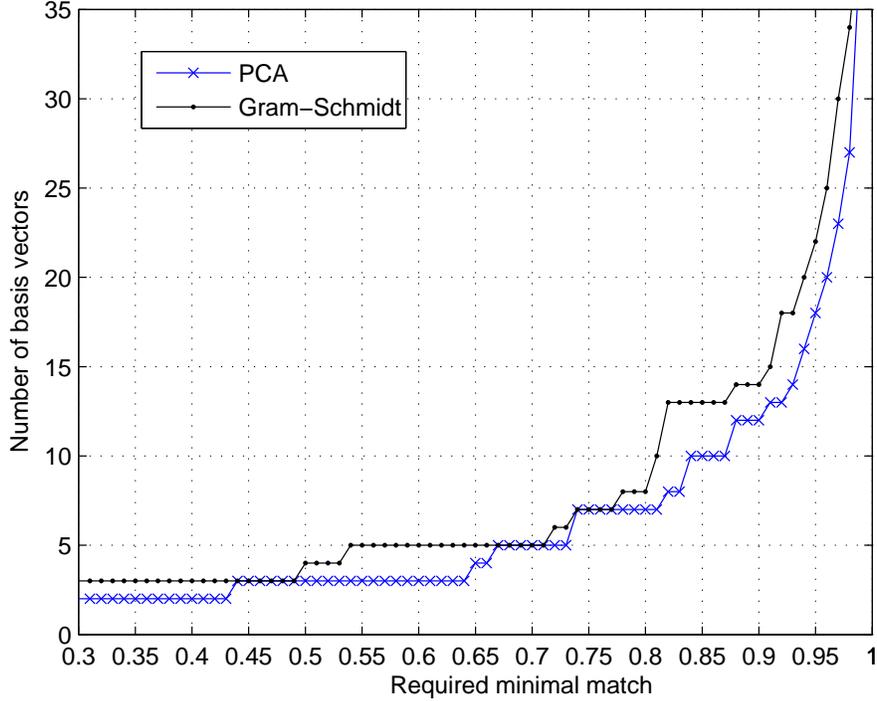, width=0.9\textwidth}
\caption{The number of PCA and GS basis vectors as a function of
minimal match. For each value of minimal match, $\mu_{min}$, we 
plot the number of basis vectors required so that $\mu >
\mu_{\rm min}$ for all waveforms in the catalogue. The number of basis
vectors required for each value of minimal match is comparable
for both methods of decomposition.
}
\label{numvecs}
\end{figure}

Computing time, however, is not a serious issue for these
waveforms because they are short and relatively few compared to, for example, 
the number of templates used in a search for gravitational waves 
from binary neutron stars (see~\cite{insp} for a recent example).
Instead, we examine the minimal match here to study the parameter space 
covered by the waveform catalogue.
If the parameter space of the waveform catalogue is degenerate, then one 
would expect the minimal match to rapidly approach 1 
for $Z \ll M$.

Figure~\ref{numvecs} shows the number of GS and PCA basis vectors
required as a function minimal match.
Similar number of GS and PCA basis vectors are required for minimal requirements up to about
0.9.
The number of basis vectors required rises rapidly as the minimal
match criterion approaches
1 since smaller features, unique to a small subset of waveforms, require a large number of
basis vectors to reconstruct.
It is interesting to note that 
for minimal match requirements greater than 0.95, more GS basis vectors are needed.
Nonetheless, the parameter space spanned by the waveform
catalogue is well spanned by less than half the total number of basis vectors for each
method.
This implies that all waveforms in the catalogue are dominated by a 
few unique features and this allows the minimal match to reach 0.75 with just 7 basis
vectors.
In fact, Dimmelmeier {\it et al.} 
noted that these waveforms can be divided into three broad categories:
waveforms due to a pressure-dominated bounce with convective overturn, waveforms
due to a pressure-dominated bounce only and waveforms with a single centrifugal 
bounce~\cite{dimmelmeier07}.

\begin{figure}[!htb]
\centering
\epsfig{file=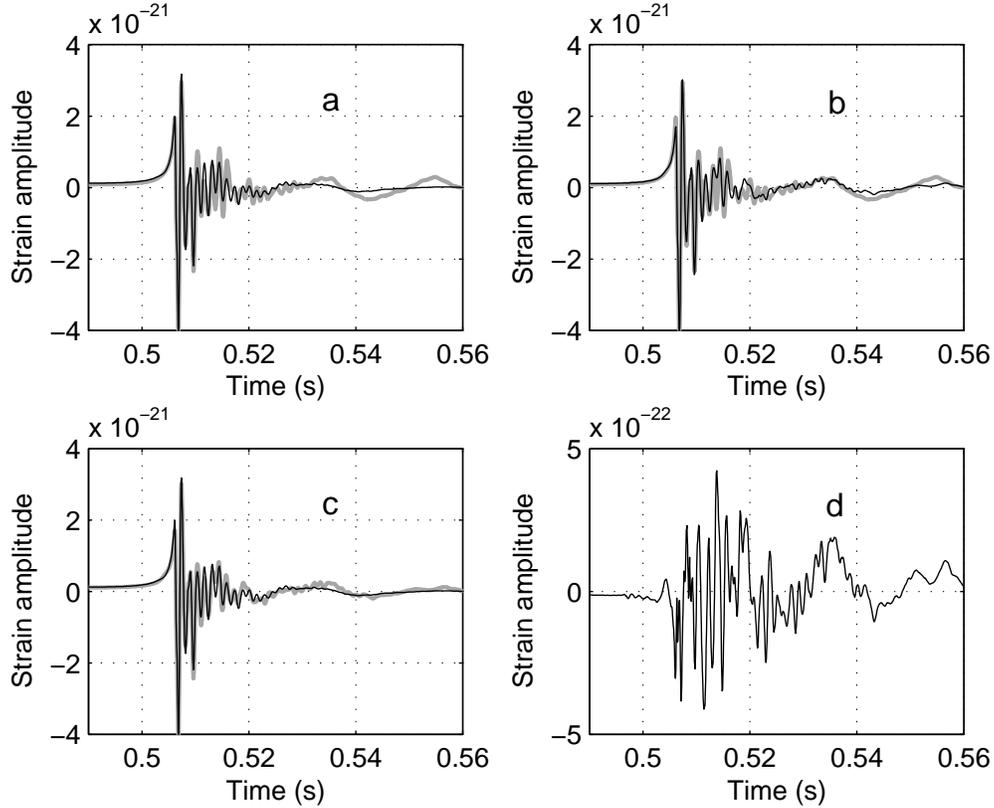, width=\textwidth}
\caption{The reconstructed waveform using PCA and GS basis vectors are
plotted as black line in figures (a) and (b) respectively. 
The grey line is the original waveform ($A=2,\beta=0.50$) from the
catalogue. 
A match of 0.95 was achieved using 6 PCA basis vectors and 17 GS 
basis vectors.
For comparison, both reconstructed waveforms are plotted in
(c), where the grey line is the PCA reconstruction and the GS
reconstruction represented by the black line. The difference
between the two reconstructions is plotted in (d).
}
\label{match_example}
\end{figure}

In Figure~\ref{match_example}, we plot an example waveform
reconstructed by the GS and PCA basis vectors with a match of
0.9. The reconstructed waveform obtained by the two sets of basis
vectors, though not identical, are very similar. The difference
between the two waveforms is only about $10\%$ in amplitude.

\subsection{Using subset of the waveforms to form basis vectors}

\begin{figure}[!htb]
\centering
\epsfig{file=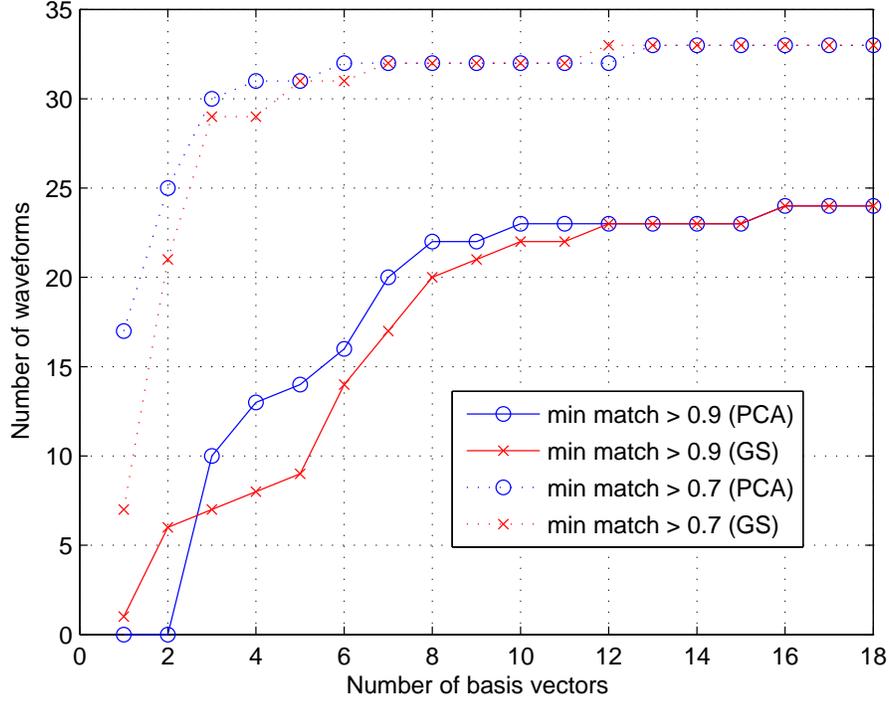, width=0.9\textwidth}
\caption{The number of waveforms with a match of at least 0.7
(dashed lines) and 0.9 (solid lines) as a function of the number
of basis vectors.
}
\label{subset}
\end{figure}

In the previous subsection, we noted that the parameter space of
the Dimmelmeier {\it et al.} waveform catalogue used in our studies can
be spanned by a small number of basis vectors.
This implies that there are many common features in the waveforms
from this catalogue. 
Here, we chose to make basis vectors using only the 18 waveforms with
moderate differential rotation at 1000\,km from the centre ($A=2$).
We make this choice to test the hypothesis that the waveforms from
precollapse stellar cores with moderate differential rotation contain features presented in
waveforms from low and highly differentially rotating stellar bodies.
Figure~\ref{subset} plots the number of waveforms with $A \neq
2$ observed to have match greater than 0.7 and 0.9.
With only 3 basis vectors, about 30 of the 36 waveforms already
have a match greater than 0.7.
With 16 PCA or GS basis vectors, $67\%$ of the remaining 36
waveforms have match greater than 0.9 and $94\%$ have match
greater than 0.7.
Therefore, a large fraction of the parameter space covered by the catalogue
is covered by waveforms from simulations with $A=2$.
This is consistent with the observations of Dimmelmeier {\it et
al}~\cite{dimmelmeier07,dimmelmeier08}
who noted that
the degree of differential rotation does not qualitatively alter the waveforms.

\section{Conclusions and discussion}

We have introduced PCA as a method of decompsing a set of waveforms into a set of basis
vectors.
A nice feature of PCA decomposition is that it allows one to
quantitatively identify the main features in a desired set of
waveforms since each basis
vector is ranked by the value of its corresponding eigenvalue.
One can interpret the basis vector with the largest corresponding eigenvalue
(the first principal component) as having the most significant features
in the waveform catalogue.

We compared the PCA method introduced here to the GS decomposition method introduced by
Brady and Ray-Majumder.
The efficiency of the PCA basis vectors at spanning the parameter space defined by the
waveform catalogue are comparable to GS decomposition, 
with about 15 basis vectors required for a minimal match of 0.9. 
For a minimal match of 0.95, 17 PCA basis vectors while 22 GS basis vectors were
required.
This shows that there are many common features in the waveforms from the chosen catalogue.
We also generated a set of basis vectors using only the 18 of the 54 waveforms (with $A=2$
only) using both methods and observed that 34 of the 36 waveforms not included in
the construction of the basis set have a minimal match of 0.7.
This implies that the features from all waveforms are well described by models with $A=2$. 

The basis vectors produced here can easily be used by parameter estimation
techniques. For example, Monte-Carlo Markov-Chain (MCMC)
methods~\cite{gilks} 
can be applied to a detected gravitational wave signal
with each basis vector as a degree of freedom to search across.
The output of the MCMC analysis would be a set of coefficients
that can be used to reconstruct the signal waveform from a linear
combination of basis vectors.
Alternatively, we can project the waveforms onto the basis
vectors to determine a set of coefficients with which we can
reconstruct each waveform with the basis vectors. 
Each waveform can then be parameterised by these coefficients
or {\it weights} and they can be used to form a
classification scheme similar to that laid out by Turk and
Pentland~\cite{eigenfaces}.
PCA as well as GS decomposition can also be used to decompose
waveforms generated by simulations from different groups, using
different core-collapse models. This application (also
proposed by Brady and Ray-Majumder~\cite{brady}) will combine the parameter
space covered by all waveforms in an efficient manner. In the
case of PCA, common features will be decomposed into the main 
components with large eigenvalues 
and, for parameter estimation, will reconstruct the main features of 
most waveforms. On the other hand, smaller features belonging to 
a small subset of waveforms will have much smaller eigenvalues 
and may be ignored by the analysis to reduce computation costs.
%

\section*{Acknowledgments}

The author would like to thank Patrick Brady, Nelson Christensen, Harald
Dimmelmeier, Martin Hendry, Christian Ott, Renate Meyer and Graham Woan for enlightening
discussions. This work is supported by the Science and Technology
Facilities Council and the Scottish Universities Physics
Alliance.


{\bf References}

\begin{thebibliography}{00}
%
\bibitem{GEO} H. L\"uck {\it et al.}, {\it Class.\ Quantum Grav.}
{\bf 23} (2006) S71--S78.
%
\bibitem{LIGO} D. Sigg (for the LIGO Scientific Collaboration),
    {\it Class.\ Quantum Grav.}\ {\bf 23} (2006) S51--6.
%
\bibitem{TAMA} M. Ando and the TAMA Collaboration,
    {\it Class.\ Quantum Grav.}\ {\bf 22} (2005) S881--9.
%
\bibitem{Virgo} F. Acernese {\it et al},
    {\it Class.\ Quantum Grav.}\ {\bf 23} (2006) S63--9.
%
\bibitem{ott-review} http://arxiv.org/abs/0809.0695
%
\bibitem{disney} M.J. Disney {\it et al.}, {\it Nature}
{\bf 455} (2008) 1082.
%
%
\bibitem{morrison} D.F. Morrison 1967, {\it Multivariate
Statistical Methods}, McGraw-Hill.
%
\bibitem{mardia} K.V. Mardia, J.T. Kent and J.M. Bibby 1980, {\it
Multivariate Analysis (Probability and Mathematical Statistics)},
Academic Press.
%
\bibitem{brady} P.R. Brady and S. Ray-Majumder, {\it Class.\
Quantum Grav.} {\bf 21} (2004) S1839--47.
%
\bibitem{zm} T. Zwerger and E. M\"{u}ller,
{\it Astron. Astrophys.} {\bf 320} (1997) 209--227.
%
\bibitem{summerscales} T.Z. Summerscales, A. Burrows, L.S. Finn
and C.D. Ott, {\it Ap. J.} {\bf 678} (2008) 1142--57.
%
\bibitem{ott2004} C.D. Ott, A. Burrows, E. Livne, and R. Walder,
 {\it Ap. J.}, {\bf 600} (2004) 834.
%
\bibitem{dimmelmeier07} H. Dimmelmeier, C.D. Ott, H.-T. Janka,
A. Marek and E. M\"{u}ller
{\it Phys. Rev. Lett.} {\bf 98} (2007) 251101.
%
\bibitem{ott-burrows} C.D. Ott, A. Burrows and Luc Dessart, {\it
Phys. Rev. Lett.} {\bf 96} (2006) 201102.
%
\bibitem{golub} G. Golub and C. van Loan 1996, {\it Matrix
Computations}, The John Hopkins University Press.
%
\bibitem{horn} R.A. Horn and C.R. Johnson 1990, {\it Matrix
Analysis}, Cambridge University Press.
%
\bibitem{SVD} G. Strang 1993, {\it Introduction to Linear
Algebra}, Wellesley-Cambridge Press.
%
%
\bibitem{shen} H. Shen, H. Toki, K. Oyamatsu and K. Sumiyoshi,
{\it Prog. Theor. Phys.} {\bf 100} (1999) 1013.
%
\bibitem{woosley} S. E. Woosley, A. Heger and T. A. Weaver, {\it Rev.
Mod. Phys.} {\bf 74} (2002) 1015.
%
\bibitem{rotation_law} H. Komatsu, Y. Eriguchi and I. Hachisu,
{\it Mon. Not. R. astr. Soc.} {\bf 237} (1989) 355--379.
%
\bibitem{insp} B. Abbott {\it et al.}, {\it Phys. Rev. D} {\bf 77} 
(2008) 062002.
%
\bibitem{dimmelmeier08} H. Dimmelmeier, C.D. Ott, A. Marek and H.-T.
Janka, {\it Phys. Rev. D} {\bf 78} (2008) 064056.
%
\bibitem{eigenfaces} M. Turk and A. Pentland,
{\it Journal of Cognitive Neuroscience} {\bf 3} (1991) 71--86
%
\bibitem{gilks} W.R. Gilks, S. Richardson and D.J. Spiegelhalter
1996, {\it Markov chain Monte Carlo in practice}, Chapman \&
Hall / CRC

\end{thebibliography}
\end{document}